\documentclass[rnote]{aa} 
\usepackage{graphicx}
\usepackage{txfonts}
\begin{document}
   \title{Evolutionary implications of the new triple-$\alpha$ nuclear reaction rate for low mass stars}


   \author{Aaron Dotter\inst{1} \and Bill Paxton\inst{2}}

   \institute{Department of Physics and Astronomy, University of Victoria,
           Victoria, BC, V8P 5C2  Canada
           \email{dotter@uvic.ca}
           \and
           Kavli Institute for Theoretical Physics,
           University of California, Santa Barbara,
           Santa Barbara, CA 93106  USA
           \email{paxton@kitp.ucsb.edu}
   }

   \date{Received \today}
 
  \abstract
   {Ogata et al. (2009; hereafter OKK) presented a theoretical determination of the 
$^4$He($\alpha\alpha,\gamma)^{12}$C, or triple-$\alpha$, nuclear reaction rate.
Their rate differs from the NACRE rate by many orders of magnitude at temperatures
relevant for low mass stars.}
   {We explore the evolutionary implications of adopting the OKK
triple-$\alpha$ reaction rate in low mass stars
and compare the results with those obtained using the NACRE rate.}
   {The triple-$\alpha$ reaction rates are compared by following the evolution of 
stellar models at 1 and 1.5 $M_{\odot}$ with Z=0.0002 and Z=0.02.}
   {Results show that the OKK rate has severe consequences
for the late stages of stellar evolution in low mass stars. Most notable is the
shortening--or disappearance--of the red giant phase.}
   {The OKK triple-$\alpha$ reaction rate is incompatible with observations 
of extended red giant branches and He burning stars in old stellar systems.}


   \maketitle
%

\section{Introduction}

OKK introduced a new determination of the triple-$\alpha$ nuclear reaction rate based
on a direct solution of the Schr\"odinger equation.
The authors provided a tabulation of the triple-$\alpha$ rate for temperatures from $10^7$ to 
$10^9$ K and also the ratio of their rate to the NACRE (Angulo et al. 1999) rate.  The OKK
rate is a factor $10^{26}$ greater than the NACRE rate at $10^7$ K, $10^6$ times greater at 
$10^8$ K, and reaches equality at $25\times10^8$ K and above.  It is important to note
that OKK normalized their rate to the NACRE result at $10^9$ K.
For comparison, the previous update to the triple-$\alpha$ reaction rate was from Fynbo et al.
(2005).  Between $10^7$ and $10^9$ K, the rate determined by Fynbo et al. differed from the NACRE
rate by a factor of $\sim3$ at most.  The previous standard rate, that of Caughlan \& Fowler (1988),
was generally lower than the NACRE rate but never by more than a factor of $\sim10$.

Such a tremendous difference at low temperatures is certain to have dramatic implications 
for stellar evolution calculations and it was with this in mind that we undertook the following 
investigation.  The purpose of this research note is simple: to investigate the evolutionary 
implications of the new triple-$\alpha$ rate and compare results obtained from the new rate with 
results based on the NACRE rate.

\section{Stellar Evolution Model Comparison}

Stellar evolution calculations were carried out using two codes: DSEP (Dotter et al., 2007)
and MESA (developed by Paxton and collaborators; http://mesa.sourceforge.net/). 
The input physics employed by DSEP and MESA
are similar for the models presented here.  Both codes use a combination of OPAL 
(Iglesias \& Rogers, 1996) and Ferguson et al. (2005) radiative opacities.  For simplicity,
the nuclear reaction networks in both codes were set to use the NACRE rates for all relevant
H- and He-burning reactions, except that either the NACRE or OKK rate was used for the 
triple-$\alpha$ process\footnote{The original version of the OKK paper contained
some typos in the tabulated rate.  The correct rate, provided to D. VandenBerg by
K. Ogata via private communication, was used in the present evolutionary sequences.}.  
For the equation of state, DSEP uses the FreeEOS code (http://freeeos.sourceforge.net)
while MESA uses a combination of OPAL (Rogers \& Nayfonov, 2002) and SCVH tables (Saumon et al. 1995)
along with the HELM EOS (Timmes \& Swesty, 2000) for temperatures and densities corresponding to He-burning
and beyond. DSEP models use a mixing length parameter of $\alpha_{MLT}$ = 1.94 and MESA models use 
$\alpha_{MLT}=2.14$. The DSEP models include gravitational settling of He and metals while the MESA models do not.

Stellar models were computed at 1 and 1.5 $M_{\odot}$ with initial X=0.70 and Z=0.02 using
both codes.  DSEP also produced models at the same masses with X=0.75, Z=0.0002.
DSEP followed the evolution from the fully-convective pre-main sequence phase until the
onset of core He burning. MESA followed the evolution from the ZAMS through the core
He flash and core He burning.  The pre-main sequence evolution is not affected by the triple-$\alpha$ process
and is not shown in the following figures for clarity.

\begin{figure*}
\centering
\includegraphics[scale=0.6]{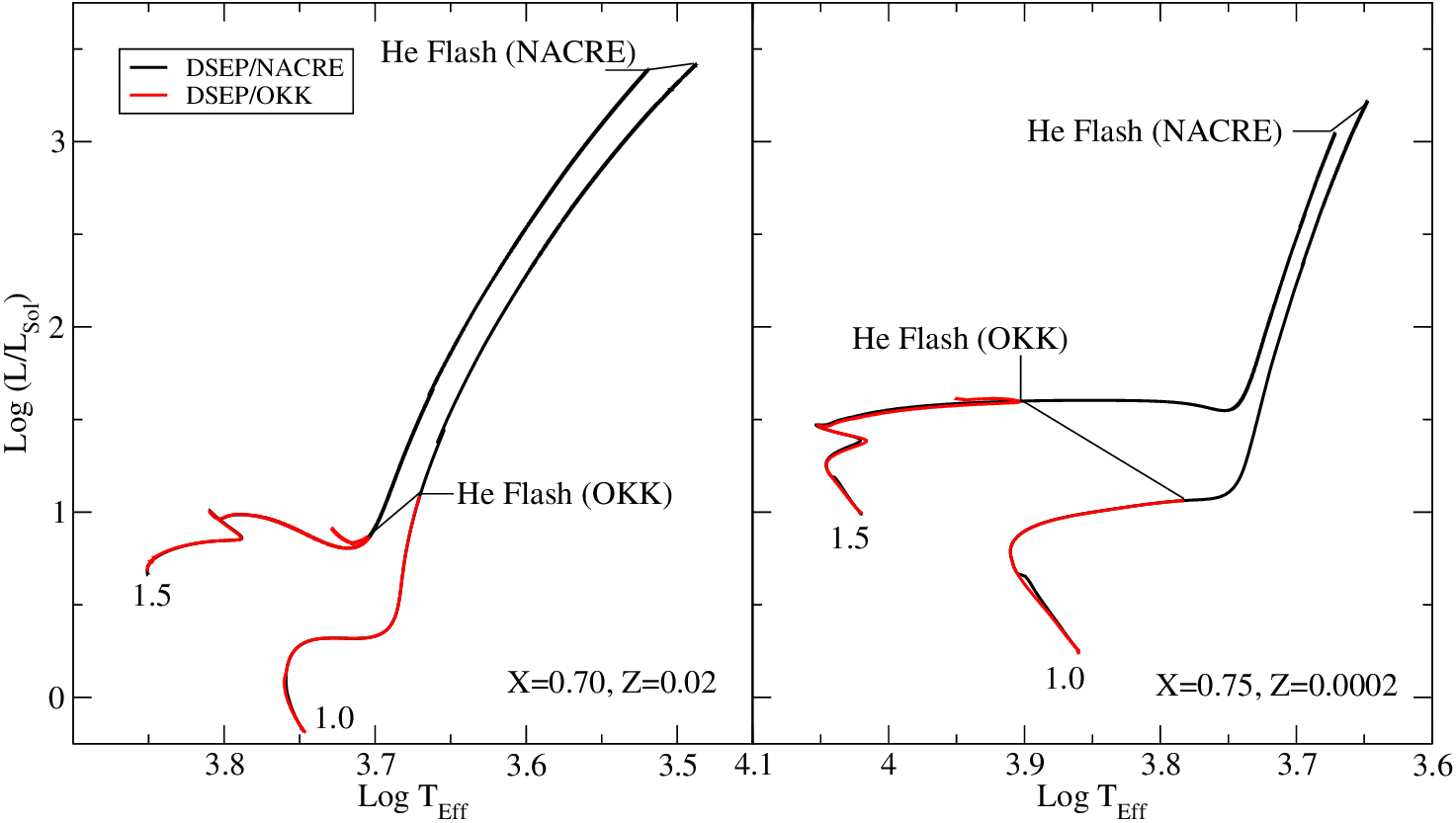}
\caption{DSEP models in the H-R diagram for X=0.7,Z=0.2 (left panel)
and X=0.75,Z=0.0002 (right panel.  Evolutionary tracks for 1 and 1.5
$M_{\odot}$ are shown in both panels.  Evolution from the ZAMS to the
onset of the core He flash is shown in the 1 $M_{\odot}$ case. For the
1.5 $M_{\odot}$ case with the OKK rate, the models transition to core He-burning 
without an explosive `flash' event and a small portion of core He-burning 
evolution, which moves the model back to the blue, is shown.}\label{DSEP}%
\end{figure*}
\begin{figure*}
\centering
\includegraphics[scale=0.6]{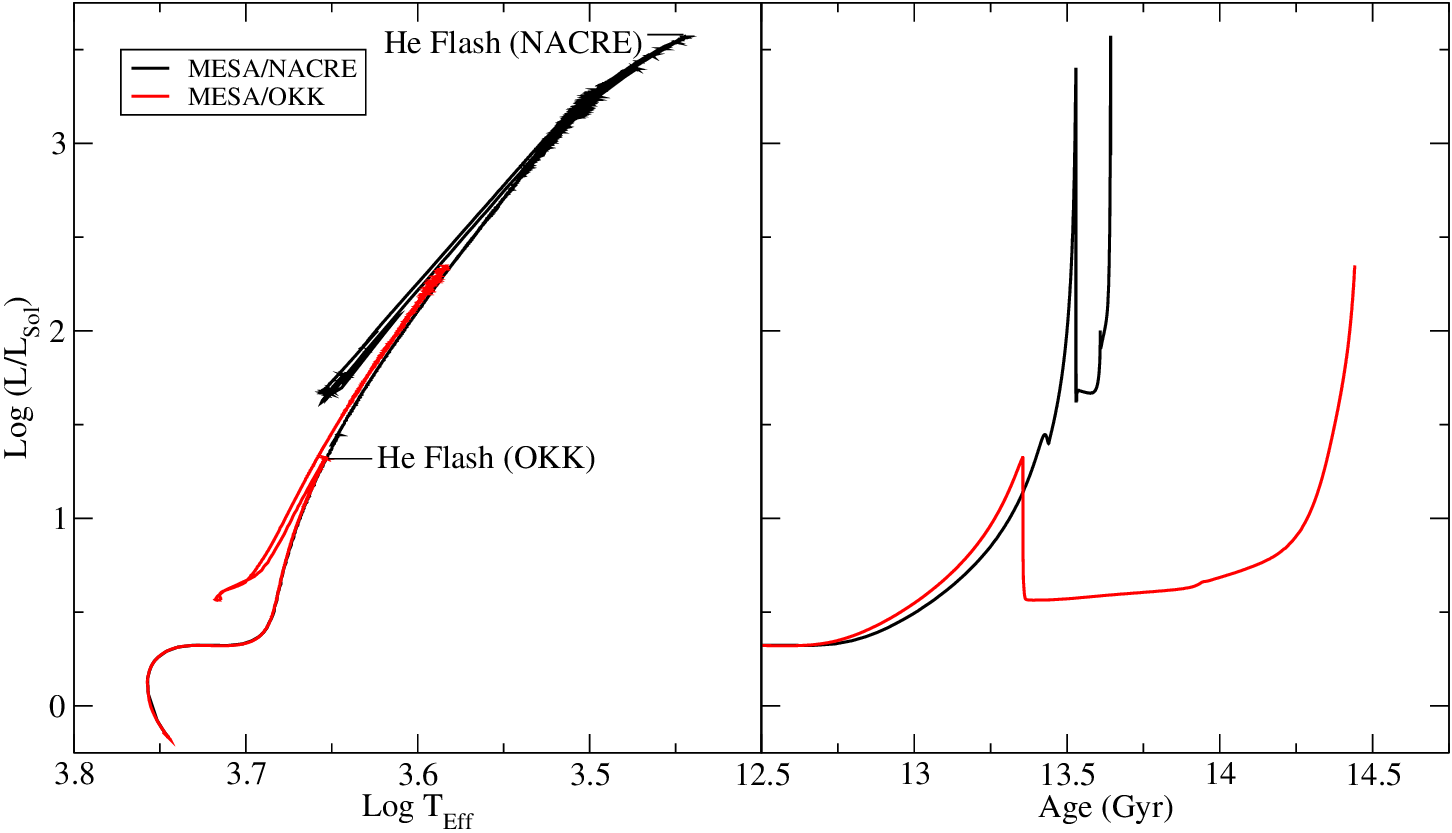}
\caption{The H-R (left panel) and age-luminosity diagrams from the 
1 $M_{\odot}$, Z=0.02 MESA model.  MESA is able to evolve the models through the 
core He flash and both tracks show core He-burning phases in the H-R diagram.  
The right panel demonstrates the prolonged core He-burning phase that results from 
adopting the OKK triple-$\alpha$ rate.}\label{MESA}%
\end{figure*}
\begin{figure*}
\centering
\includegraphics[scale=0.6]{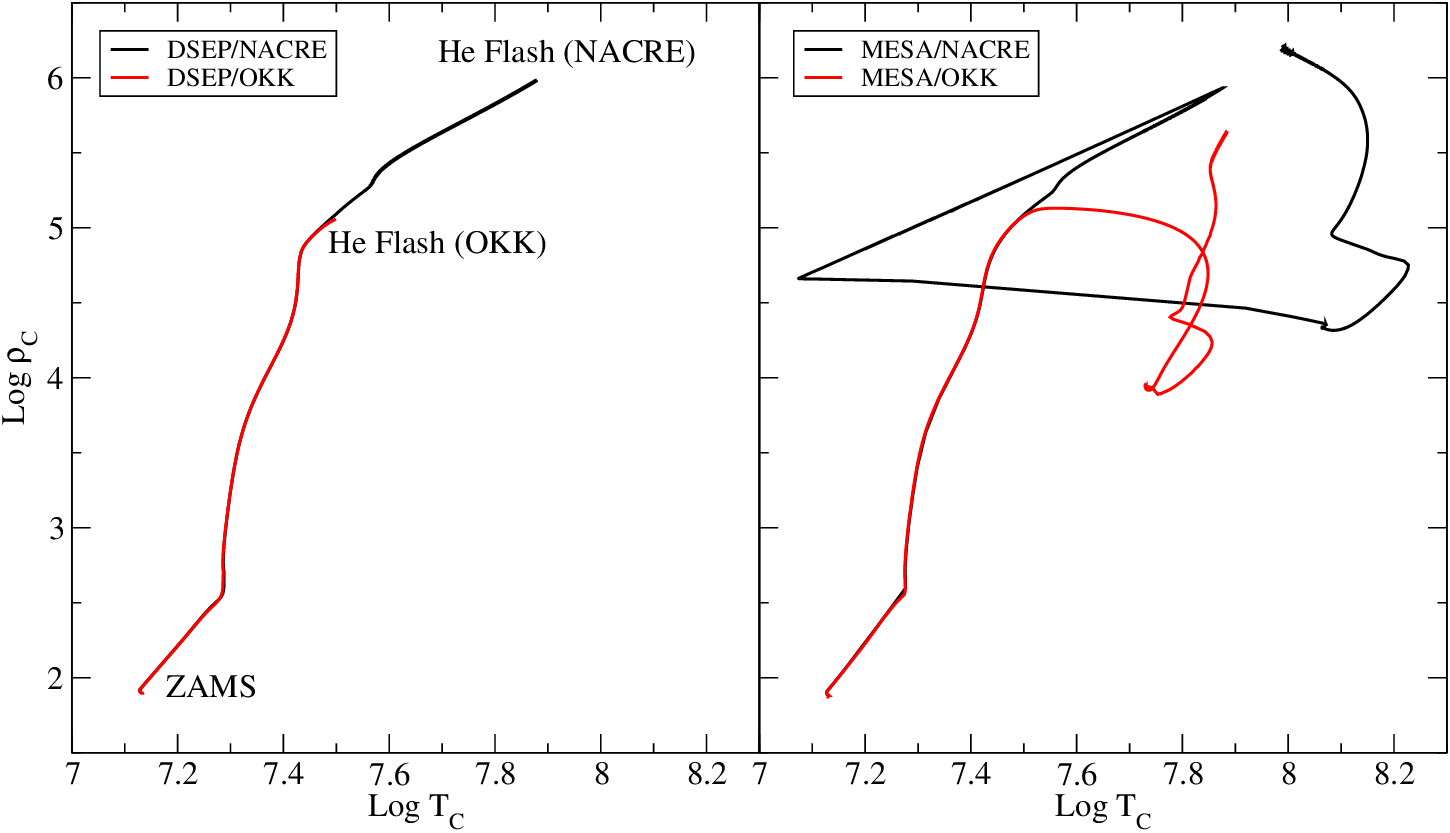}
\caption{Central temperature-density diagram for 1 $M_{\odot}$ models
at Z=0.02.  The left panel shows the DSEP models and labels the location
of the ZAMS and core He flash for both models.  The right panel shows MESA
models, which also show the core He-burning phase of evolution.}\label{central}%
\end{figure*}

Figure \ref{DSEP} shows the H-R diagram with DSEP models at 1 and 1.5 $M_{\odot}$:
the left panel includes tracks at X=0.7,Z=0.02 and the right panel shows tracks
at X=0.75,Z=0.0002.  Both panels clearly demonstrate that the DSEP/NACRE models
experience an extended red giant phase in all cases.  Conversely, the DSEP/OKK 
models show little or no red giant phase, with the Z=0.0002 models transitioning
to core He-burning while still in the subgiant phase. The OKK rate also causes a 
very modest increase in surface temperature during the main sequence but the effect 
is small and difficult to see in the figure.

Figure \ref{MESA} shows, in the left panel, the H-R diagram of the 1 $M_{\odot}$, Z=0.02 
model evolved through the core He flash and core He-burning phase by the MESA code.  
The MESA/OKK model has a core
He-burning phase that occurs at $\sim$3-4 $L_{\odot}$: about an order of magnitude
fainter than the core He-burning phase experienced by MESA/NACRE model.  The right
panel of Figure \ref{MESA} shows the late-time evolution of the same models.  
Because the MESA/OKK model undergoes core He-burning at such low luminosity,
it is able to continue for about 1 Gyr at more or less constant luminosity.  The
MESA/NACRE model experiences core He-burning for $\sim$100 Myr at roughly ten times
the luminosity of the MESA/OKK model.

Figure \ref{central} shows the central temperature-density diagrams for the 1 
$M_{\odot}$, Z=0.02 models with DSEP on the left and MESA on the right.  Models
evolved with both codes trace out identical paths in this plane until the
DSEP models terminate at the onset of the core He flash.  The MESA tracks
show the evolution through the core He flash and core He-burning phases.
The figure demonstrates that the OKK rate causes the core He flash to occur at a 
much lower temperature ($\sim$30 million K  vs. $\sim$80 million K) and an order 
of magnitude lower density than the NACRE rate.

\section{Conclusions}

Comparisons of stellar evolution models computed with NACRE and OKK triple-$\alpha$
reaction rates were presented.  The OKK rates cause models of low mass stars to have
either a shortened red giant phase or to bypass the red giant phase altogether. The
OKK rate also causes core He-burning to take place at a factor of ten lower
luminosity--and for a factor ten longer duration--than the NACRE rate. 

Given the excellent agreement between stellar evolution models and observations of
the RGB (Salaris, Cassisi, \& Weiss, 2002) when the models employ the NACRE (or relatively similar)
triple-$\alpha$ reaction rates,  such as Caughlan \& Fowler (1988) or Fynbo et al. (2005), it is difficult 
to envision how such a dramatic change as that proposed by OKK could lead to improved agreement with 
observations.  Indeed, the triple-$\alpha$ reaction rate determined by OKK is in stark 
disagreement with fundamental observational evidence of stellar evolution, namely: (1) 
the existence of extended red giant branches in old stellar systems, (2) the location of 
core He-burning stars in the H-R or color-magnitude diagram, and (3) the lifetime ratio of 
the core He-burning phase to the red giant phase as manifested by the number ratio 
of horizontal branch stars to RGB stars, also known as the R parameter 
(e.g. Salaris et al. 2005).

\begin{acknowledgements}
AD was supported by a CITA National Fellowship and by an NSERC grant to D. VandenBerg.
BP was supported by the National Science Foundation under grants PHY 05-51164 and AST 07-07633.
\end{acknowledgements}

\end{document}